\documentclass{ws-p9-75x6-50}

\begin{document}

\title
{
Electronic phase diagram of La$_{1.875}$Ba$_{0.125-x}$Sr$_x$CuO$_4$
}

\author{M. Fujita, H. Goka, K. Yamada}

\address{Institute for Chemical Research, Kyoto University, Gokasyo, Uji 611-0011, Japan\\E-mail: fujita@scl.kyoto-u.ac.jp}


\maketitle

\abstracts{
We performed systematic measurements of magnetic susceptibility on single crystals of La$_{1.875}$Ba$_{0.125-x}$Sr$_x$CuO$_4$. The dependence of the superconducting transition temperature on Sr-concentration demonstrates a step-like pattern upon doping at {\it x}$\sim$0.08 as the crystal structure changes from low-temperature tetragonal (LTT) to low-temperature orthorhombic (LTO) phase at low temperature. Upon cooling, an anomalous upturn in the susceptibility was observed at the  structural phase transition  between the LTT-LTO phases under the magnetic field parallel to {\it c}-axis.
}

\section{Introduction}

The discovery of anomalous suppression of superconductivity in La$_{2-x}$Ba$_x$CuO$_4$ system with {\it x}$\sim$1/8~\cite{Moodenbaugh88}~\cite{Kumagai88}, an issue dubbed the 1/8 problem, has been paid remarkable attention in the field superconductivity. In this system, a structural phase transition from the low temperature orthorhombic (LTO) to the low temperature tetragonal (LTT) phase occurs in a narrow range of Ba concentrations surrounding 1/8.~\cite{Axe89} A similar suppression of superconductivity is also observed in La$_{2-y-x}$Nd$_y$Sr$_x$CuO$_4$ in the LTT phase around {\it x}=1/8.~\cite{Crawford91} The superconductivity of La$_{2-x}$Sr$_x$CuO$_4$ in the LTO phase, however, is suppressed around {\it x}=0.115.~\cite{Takagi89}~\cite{Kumagai94} Thus, suppression of superconductivity is a generic feature of the hole-doped La-214 system, dependant on crystal structure.

The recent discovery of both spin and charge stripe orders in the LTT phase of the La$_{1.6-x}$Nd$_{0.4}$Sr$_x$CuO$_4$ system reveald a new aspect of the 1/8 problem.~\cite{Tranquada95} A sample with {\it x}$\approx$1/8 possesses the highest stripe ordering temperature and the lowest superconducting transition temperature ({\it T}$_c$), suggesting a competition between stripe order and superconductivity.~\cite{Tranquada97} Moreover, since the stripe ordering develops below the LTO-LTT transition temperature ({\it T}$_{d2}$), the LTT lattice potential is favorable for the pinning of dynamical fluctuations of spin/charge stripes. The static/quasi-static magnetic order, however, is also observed in orthorhombic La$_{2-x}$Sr$_x$CuO$_4$.~\cite{Kumagai94}~\cite{Goto94}~\cite{Kimura99} These results suggest that stripe order is an inherent property of the La-214 system at {\it x}$\approx$1/8, irrespective of crystal structure, whose stability is affected by the lattice potential.

%Fig1========================================================
\begin{figure}[t]
\centerline{\epsfxsize=5in\epsfbox{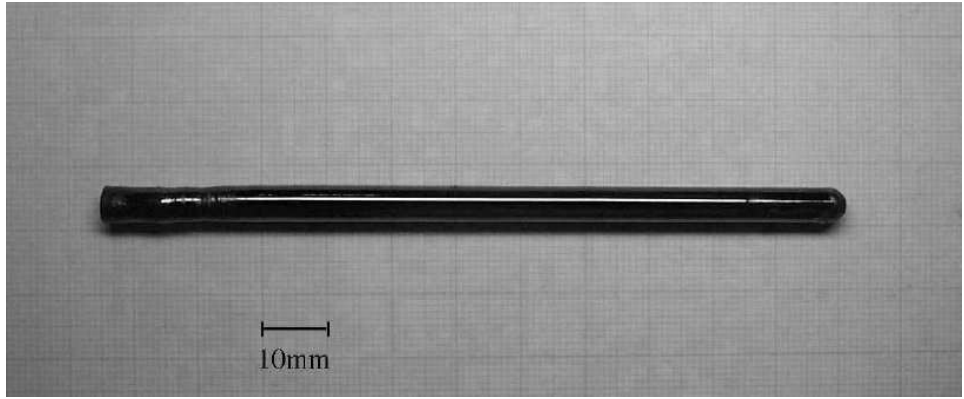}}
\caption{Single crystal rod of La$_{1.875}$Ba$_{0.04}$Sr$_{0.085}$CuO$_4$}
\end{figure}
%============================================================

To clarify the relationship between the suppression of superconductivity, stripe order and crystal structure, we investigated the electronic phase diagram of the La$_{1.875}$Ba$_{0.125-x}$Sr$_x$CuO$_4$ (LBSCO) system. At low temperature, this system allows the crystal structure to change from LTT to LTO through the variation of the Ba/Sr doping ratio while keeping the total hole concentration of 1/8 constant.~\cite{Maeno91} LBSCO is a suitable compound to investigate the physical properties of the 1/8 problem without the effects of a large, rare-earth moment.
In this paper, we measured the magnetic susceptibility of LBSCO single crystals at varying several doping levels. We discovered a clear $x$-dependence of {\it T}$_c$ and clarified an anomalous behavior of the susceptibility.

\section{Experimental Details}
Single crystals of LBSCO with {\it x}=0.05, 0.06, 0.075, 0.085 and 0.1 were grown using a traveling-solvent floating-zone method with two infrared furnaces (Nichiden Machinery, SC-N35HD-E and SC-K15HD) Growth conditions were similar to those used to generate a Ba-free La$_{2-x}$Sr$_{x}$CuO$_4$ single crystal.~\cite{Hosoya94} Large single crystals with high Ba concentrations are lacking due to difficulties in growth. We have produced single crystal rods with the typical diameter of $\sim$6mm and a length of $\sim$80mm by using large, focusing mirrors. Fig.1 exhibits a single crystal rod of La$_{1.875}$Ba$_{0.04}$Sr$_{0.085}$CuO$_4$ $\sim$100mm in length. A longer crystal rod reduces the concentration gradient in the direction of growth. Magnetic susceptibility measurements on several parts of a crystal rod demonstrate the saturation of {\it T}$_c$ at growth lengths greater than $\sim$50mm. Samples were annealed under oxygen gas flow for 50 hours at  900$^{\circ}${\it C} and then cooled to 500$^{\circ}${\it C} at a rate of 10$^{\circ}${\it C}/h. Following a subsequent annealing at 500$^{\circ}${\it C} for 50 hours, samples were subjected to furnace-cooling to reduced them to room temperature. 
To determine {\it T}$_c$ (onset), we measured the diamagnetic susceptibility at the final part of growth by using a SQUID magnetometer(Quantum Design, MPMS 2 and MPMS XL) under a magnetic field of 10 Oe following the zero-field-cooling process. In a magnetic field of 50000 Oe, parallel to a$_{tetra}$- and c-axes, we measured the temperature dependence of susceptibility for a single crystals of La$_{1.875}$Ba$_{0.075}$Sr$_{0.05}$CuO$_4$.

%Fig2========================================================
\begin{figure}[h]
\centerline{\epsfxsize=4.9in\epsfbox{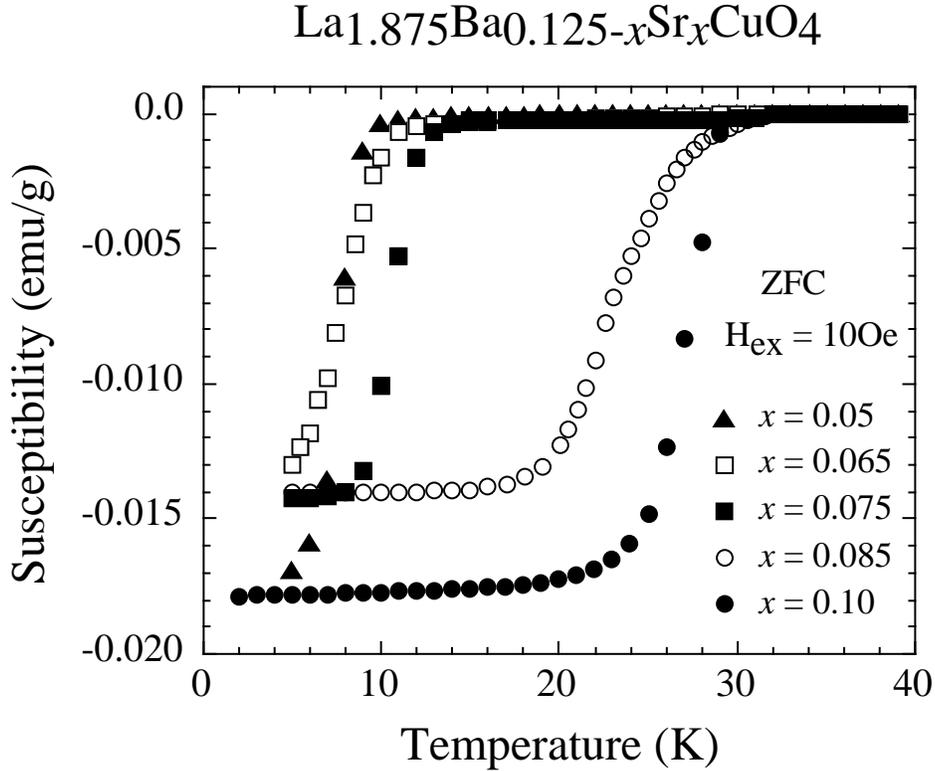}}
\caption{Magnetic susceptibility measured at 10Oe for zero-field-cooled single crystals of La$_{1.875}$Ba$_{0.125-x}$Sr$_x$CuO$_4$ with  {\it x}=0.05, 0.06, 0.075, 0.085 and 0.1.}
\end{figure}
%============================================================

\section{Results and Discussion}

We measured the temperature dependence of diamagnetic susceptibility for a series of crystals(Fig.2). Samples with {\it x}=0.05, 0.06, 0.075, 0.085 and 0.1 exhibit a superconducting transition at {\it T}$_c$=10K, 11.5K, 14K, 32K and 30.5K, respectively. {\it T}$_c$ changes discontinuously, however, upon Sr-doping at a concentration of approximately  {\it x}=0.08; {\it T}$_c$ is $\sim$30K for {\it x}$\geq$0.085 and is $\sim$12K for {\it x}$\leq$0.075 . The magnetic susceptibilities for {\it x}$\leq$0.075 exhibit an anomaly around 30K, corresponding to the superconducting transition of the residual LTO phase. Such a residual LTO phase is also reported by a recent, high-resolution, neutron power diffraction study examining LBSCO. ~\cite{Lappas00}

In Fig.3, we summarize the {\it T}$_c$ of 1/8-doped LBSCO system  as a function of Sr concentration {\it x}. The {\it T}$_c$ of Ba-free La$_{2-x}$Sr$_{x}$CuO$_4$ with either $x$=0.12 or 0.13 are listed as a reference.~\cite{Kimura99}~\cite{Matsushita99} As {\it x} increases, {\it T}$_c$ rises slowly from {\it x}$\leq$0.075, then increases rapidly around {\it x}=0.08, saturating around {\it T}=30K for {\it x}$\geq$0.085. By plotting  {\it T}$_{d2}$ as a function of {\it x} (the dashed line in Fig.3)~\cite{Maeno91}, a close relation between highly reduced,
\linebreak
%Fig3========================================================
\begin{figure}[t]
\centerline{\epsfxsize=4.6in\epsfbox{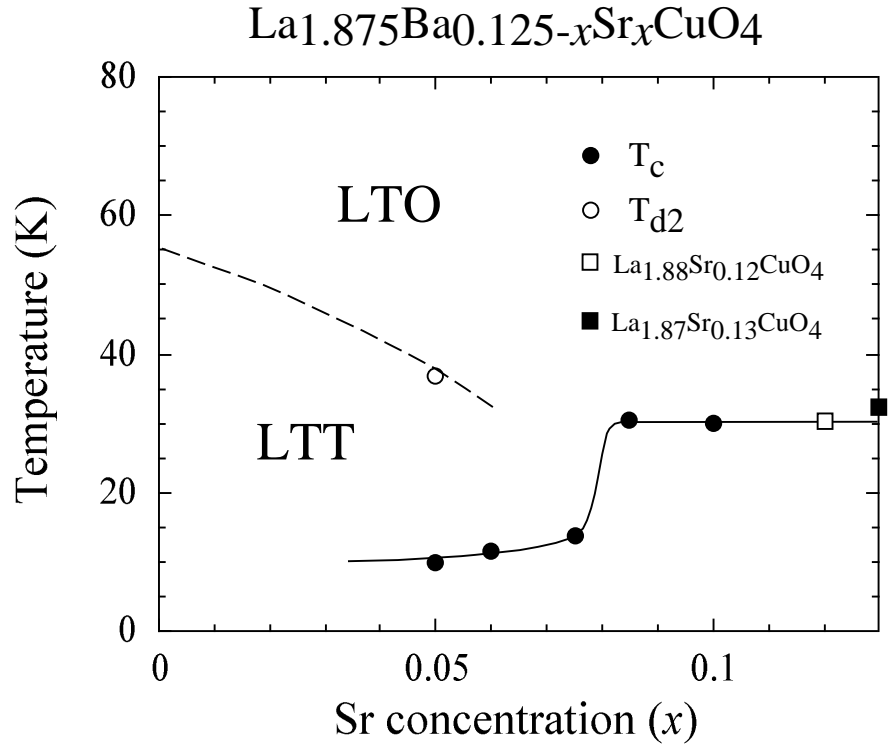}}
\caption{Phase diagram of superconducting transition temperatures ({\it T}$_c$) for La$_{1.875}$Ba$_{0.125-x}$Sr$_x$CuO$_4$ as a function of {\it x}(closed circles) and the LTO-LTT transition temperature ({\it T}$_{d2}$) as determined by X-ray diffraction(dashed line).$^{ 12}$ The {\it T}$_c$ of Ba-free La$_{2-x}$Sr$_{x}$CuO$_4$ with either $x$=0.12(open square) or 0.13(closed square) is listed.$^{ 11}$ $^{ 16}$
The open circle represent the {\it T}$_{d2}$ of a sample possessing an {\it x}=0.05 as  determined by neutron scattering measurement.$^{ 17}$ The solid line demonstrates the overall shape of the {\it T}$_{d2}$ function.}
\end{figure}
%============================================================
\noindent
depending on {\it T}$_{d2}$ as compared to the LTO phase in which {\it T}$_{c}$ is independent of Ba/Sr ratio. This phenomenon is more clearly observed in our present measurements using single crystals compared to previous studies utilizing powder samples.

The graph of {\it T}$_c$ as a function of $x$ reveals a first-order transition-like behavior. In contrast, {\it T}$_{d2}$ as determined by powder X-ray measurement, demonstrates a gradual change similar to a second-order transition upon doping. A close relationship between crystal structure and suppression of superconductivity should produce a sharp transition in the $x$ dependence of {\it T}$_{d2}$ at approximately $x$=0.08. Thus, the $x$-dependence of {\it T}$_{d2}$ using single crystals should be examined in more detail.

%Fig4========================================================
\begin{figure}[t]
\centerline{\epsfxsize=5.1in\epsfbox{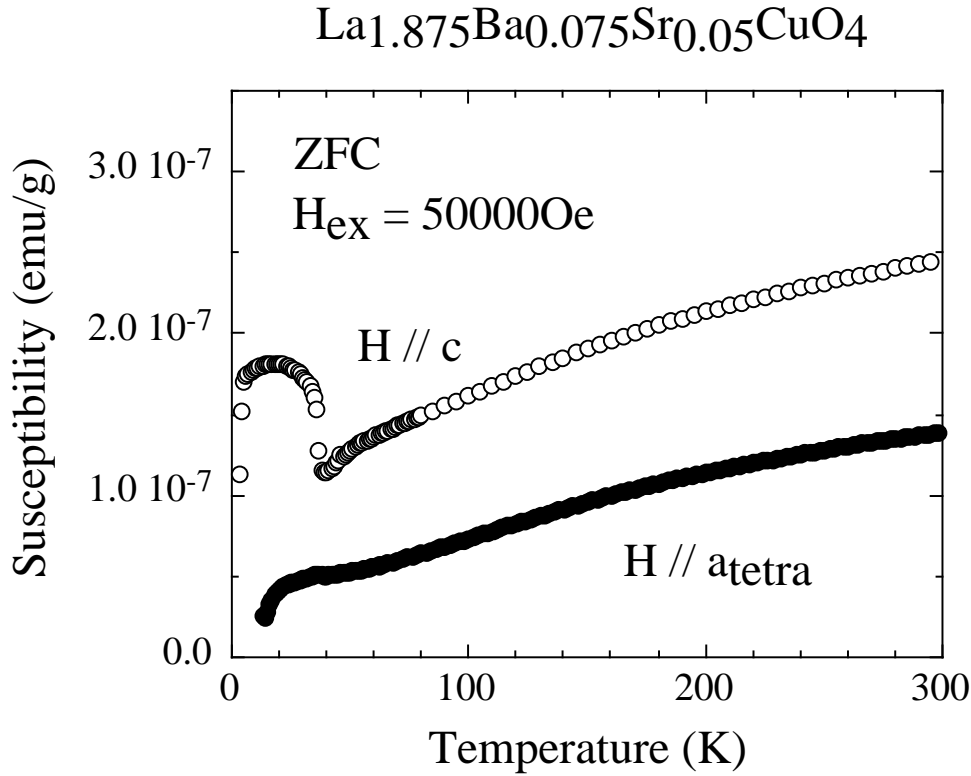}}
\caption{Magnetic susceptibility of La$_{1.875}$Ba$_{0.075}$Sr$_{0.05}$CuO$_4$. Magnetic field of 50000 Oe was applied along the a$_{tetra}$-(closed circles) and c-axes(open circles).}
\end{figure}
%============================================================

We measured the magnetic susceptibility of La$_{1.875}$Ba$_{0.075}$Sr$_{0.05}$CuO$_4$ measured under a magnetic field of 50000 Oe parallel to the a$_{tetra}$- or c-axes(Fig.4). Under the latter field, an anomalous upturn is observed at $\sim$38K as the temperature is lowered. This anomalous behavior is independent of the applied field amplitude. The upturn occurs at {\it T}$_{d2}$, suggesting a simultaneous change in both the spin and the crystal structure, as discussed with high field MNR experiments on La$_{1.885}$Sr$_{0.115}$CuO$_4$.~\cite{Goto97} It is necessary to perform a systematic study of LBSCO to understand the interplay between the anomalous behavior altering magnetic susceptibility, the LTT-LTO transition, and the suppression of superconductivity.

\section*{Acknowledgements}
We would like to thank G.Shirane, H. Kimura and J.M.Tranquada for stimulating discussions. This work was supported by a Grand-In-Aid for Scientific Research from Japanese Ministry of Education, Science, Sports and Culture.


\begin{thebibliography}{99}

\bibitem{Moodenbaugh88} A. R. Moodenbaugh, Y. Xu. M.Suenaga, T. J. Folkerts, and R. N. Shelton, {\em Phys. Lev.} B {\bf 38}, 4596 (1988)

\bibitem{Kumagai88} K. Kumagai, Y. Nakamua, I. Watanabe, Y. Nakamichi, and H. Nakajima, {\em J. Mag, Mag, Mater.} {\bf 76-77}, 601 (1988)

\bibitem{Axe89} J. D. Axe, A. H. Moudden, D. Hohlwein, D. e. Cox, K. M. Mohanty, A. R. Moodenbaugh, and Y. Xu, {\em Phys. Rev. Lett.} {\bf 62}, 2751 (1989)

\bibitem{Crawford91} M. K. Crawford, R. L. Harlow, E. M. MaCarron, W. E. Farneth, J. D. Axe, H. Chou and Q. Huang, {\em Phys. Rev.} B {\bf 44}, 7749 (1991)

\bibitem{Takagi89} H. Takagi, T. Ido, S. Ishibashi, M. Uota, S. Uchida, and Y. Tokura, {\em Phys. Rev.} B. {\bf 40}, 2254 (1989)

\bibitem{Kumagai94} K. Kumagai, K. Kawano, I. Watanabe, K. Nishiyama, and K. Nagamine, {\em J. Supercond.} {\bf 7}, 63 (1994)

\bibitem{Goto94} T. Goto, S. Kazama, K. Miyagawa and T. Fukase, {\em J. Phys. Soc. Jpn.} {\bf 63}, 3494 (1994)

\bibitem{Tranquada95} J. M. Tranquada, B. J. Sternlieb, J. D. Axe, Y. Nakamura, and S. Uchida, {\em Nature} (London) {\bf 375}, 561 (1995)

\bibitem{Tranquada97} J. M. Tranquada, J. D. Axe, N. Ichikawa, A. R. Moodenbaugh, Y. Nakamura, and S. Uchida, {\em Phys. Rev. Lett.} {\bf 78}, 338 (1997)

\bibitem{Kimura99} H. Kimura, K. Hirota, H. Matsushita, K. Yamada, Y. Endoh, S. -H. Lee, C. F. Majkrzak, R. Erwin, G. Shirane, M. Greven, Y. S. Lee, M. A. Kastner, and R. J. Birgeneau, {\em Phys. Rev.} B {\bf 59}, 6517 (1999).

\bibitem{Maeno91} Y. Maeno, A. Odagawa, N. Kakehi, T. Suzuki, and T. Fujita, {\em Physica} C {\bf 173} (1991) 322 

\bibitem{Hosoya94} S. Hosoya, C. H. Lee, S. Wakimoto, K. Yamada, and Y. Endoh, {\em Physica} C {\bf 235-240}, 547 (1994).

\bibitem{Katano95} S. Katano, Y. Ueda, A. Hayashi, N. M$\hat{o}$ri, {\em Physica} B {\bf 213-214} (1995) 81

\bibitem{Lappas00} A. Lappas, K. Prassides, F. N. Gygax, and A. Schenck, {\em J. Phys.} {\bf 12}, 3401 (2000).

\bibitem{Matsushita99} H. Matsushita, H. Kimura, M. Fujita, K. Yamada, K. Hirota, Y. Endoh, {\em J. Phys. Chem. Solid.} {\bf 60}, 1071 (1999).

\bibitem{Fujita00} M. Fujita, H. Goka, K. Yamada, unpublished data

\bibitem{Goto97} T. Goto, K. Chiba, M. Mori, T. Suzuki, and T. Fukase, {\em J. Phys. Soc. Jpn.} {\bf 66}, 2870 (1997)


\end{thebibliography}
\end{document}